\documentclass[aps,prl,reprint,superscriptaddress,showpacs]{revtex4-1}
\usepackage{graphicx}
\usepackage{dcolumn}
\usepackage{bm}
 \usepackage{tensor}
 \usepackage{physics}
\usepackage{times}
\usepackage{siunitx}

\begin{document}


\title{\vspace{-0.5 cm} Sorting photons by radial quantum number}


\author{Yiyu Zhou}
\affiliation{The Institute of Optics, University of Rochester, Rochester, New York 14627, USA}
\author{Mohammad Mirhosseini}
 \email{mirhosse@optics.rochester.edu}
  \altaffiliation[Current address: ]{Thomas J. Watson, Sr., Laboratory of Applied Physics,
California Institute of Technology, Pasadena, California 91125, USA}
\affiliation{The Institute of Optics, University of Rochester, Rochester, New York 14627, USA}
\author{Dongzhi Fu}
\affiliation{The Institute of Optics, University of Rochester, Rochester, New York 14627, USA}
\affiliation{Key Laboratory for Quantum Information and Quantum Optoelectronic Devices, Department of Applied Physics, Xi'an Jiaotong University, Xi'an, Shaanxi Province, 710049, China}
\author{Jiapeng Zhao}
\affiliation{The Institute of Optics, University of Rochester, Rochester, New York 14627, USA}
\author{\\Seyed Mohammad Hashemi Rafsanjani}
\affiliation{The Institute of Optics, University of Rochester, Rochester, New York 14627, USA}
\author{Alan E. Willner}
 \affiliation{Department of Electrical Engineering, University of Southern California, Los Angeles, CA 90089, USA}
\author{Robert W. Boyd}
\affiliation{The Institute of Optics, University of Rochester, Rochester, New York 14627, USA}
\affiliation{Department of Physics, University of Ottawa, Ottawa, Ontario K1N 6N5, Canada}



\date{\today}

\begin{abstract}
The Laguerre-Gaussian (LG) modes constitute a complete basis set for representing the transverse structure of a {paraxial} photon field in free space. Earlier workers have shown how to construct a device for sorting a photon according to its azimuthal LG mode index, which describes the orbital angular momentum (OAM) carried by the field. In this paper we propose and demonstrate a mode sorter based on the fractional Fourier transform (FRFT) to efficiently decompose the optical field according to its radial profile. We experimentally characterize the performance of our implementation by separating individual radial modes as well as superposition states. The reported scheme can, in principle, achieve unit efficiency and thus can be suitable for applications that involve quantum states of  light. This approach can be readily combined with existing OAM mode sorters to provide a complete characterization of the transverse profile of the optical field.
\vspace{-0.5 cm}
\end{abstract}


\maketitle

In the recent years, the transverse structure of optical photons has been established as a resource for storing and communicating quantum information \cite{mair2001entanglement}. In contrast to the two-dimensional Hilbert space of polarization, it takes an unbounded Hilbert space to provide a mathematical representation for the transverse structure of the optical field. The large information capacity of structured photons has been recently utilized to enhance quantum key distribution (QKD) \cite{Groblacher2006,Walborn:2006jv,Mafu:2013vr,mirhosseini2015high} and a multitude of other applications \cite{franke2008advances, nagali2009quantum, wang2012terabit, Bozinovic:2013ch, Krenn29042014}.
The orbital angular momentum (OAM) modes have become increasingly popular for implementing multi-dimensional quantum states due to the relative ease in generation \cite{Gibson2004}, manipulation \cite{malik2016multi}, and characterization of these modes \cite{berkhout2010efficient, mirhosseini2016wigner}.

Although the OAM modes provide a basis set for representing the azimuthal structure of photons, they cannot completely span the entire transverse state space, which encompasses an extra (radial) degree of freedom. The Laguerre-Gaussian (LG) mode functions provide a basis to fully represent the spatial structure of the transverse field \cite{simpson1996optical, allen1992orbital, salakhutdinov2012full}. These modes are characterized by two numbers, the radial mode index $p \in \{0,1,2,...\}$ and the azimuthal mode index $\ell \in \{0,\pm 1,\pm 2,...\}$. While the azimuthal number $\ell$ is well studied due to its association with the OAM of light \cite{allen1992orbital}, the radial index $p$ has so far remained relatively unexplored. {The quantum coherence of photons in a superposition of orthogonal radial modes has been recently demonstrated in the context of quantum communication and high-dimensional entanglement \cite{karimi2014exploring, krenn2014generation, salakhutdinov2012full}.} The radial LG modes also hold a number of promising features, and have been studied in the contexts of self-healing \cite{mendoza2015laguerre}, superresolution \cite{ tsang2016quantum}, and hyperbolic momentum charge \cite{plick2015physical}. Despite the growing theoretical interests in utilizing the radial structure of photons, the experimental realizations have thus far been impeded because of the difficulty of measuring these modes.

The initial step in characterization of the radial degree of freedom of light is to find a radial mode spectrum, i.e. to find the probability $P(p)$ of having the state prepared in mode index $p$. This information can be, in principle, obtained by performing a series of projective measurements. However, the most straightforward method for implementing the projective measurement of a radial LG mode requires shaping the amplitude of the incoming light beam, and the resulting loss makes this approach unsuitable for operation at the single-photon level \cite{qassim2014limitations}. In addition to this technical difficulty, the projective measurement of a photon results in its absorption \cite{Krenn29042014}. This inherently limits the success rate to $1/d$ in a $d$-dimensional state space, a rate that does not scale well with the size of the Hilbert space. An alternative approach for characterizing the radial mode structure is to sort an unknown incoming photon by its radial quantum number. A radial mode sorter would route the photon to a distinct output that is indexed by the value of its radial quantum number $p$, and is thus capable of performing parallel projective measurements with a success rate of unity.

Here, {we propose and demonstrate a unitary mode sorter for the radial quantum number $p$.} Our approach relies on a key property of the Laguerre-Gaussian modes: the dependence of the effective phase velocity on the radial quantum number $p$. We use a set of refractive optical elements to induce the fractional Gouy phase by realizing a fractional Fourier transform (FRFT) module \cite{yew2008fractional}. The FRFT module is then combined with a Mach-Zehnder interferometer that can discriminate the modes based on the magnitude of the induced phase. Our experiment can be understood as an implementation of the theoretical recipe recently developed in Ref.\,\cite{ionicioiu2016sorting}. We provide experimental results demonstrating the ability to sort individual and superposition states residing in the 4-dimensional state space of $p \in \{0, 1, 2, 3 \}$. Furthermore, we show that our implementation can be combined with the existing methods of sorting OAM to provide full characterization of the transverse structure of the light field.

\begin{figure*}[t]
\includegraphics[width=  \textwidth]{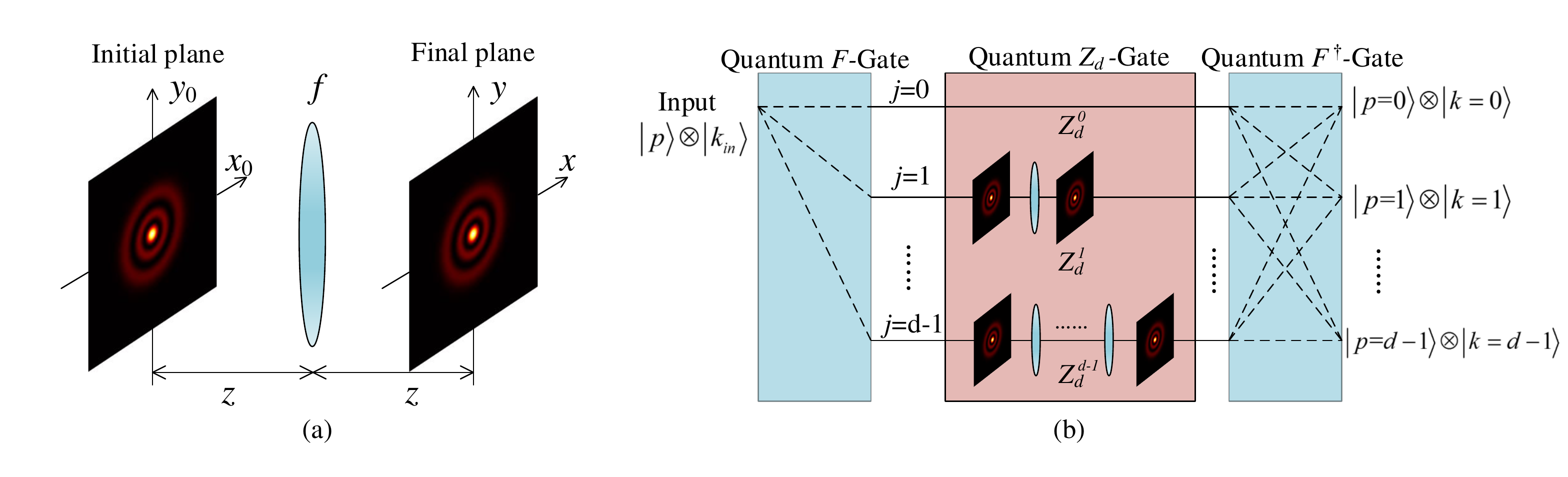}
\caption{
(a). Realization of the fractional Fourier transformation (FRFT) with a single lens. The Laguerre-Gaussian functions are the eigenmodes of the FRFT and thus maintain their shape under this transformation. Here a $p=2$ mode is shown as an example. (b). A $d$-dimensional quantum sorter composed of a discrete $F$-gates and a $Z_d$-gate. The $Z_d$-gate is implemented by the FRFT in our experiment. Note that the design can be simplified by replacing the first $F$-gate with an 1-to-$d$ beam splitter, which is permissible because the system has only one effective input port.
 }
\label{fig:FRFT}
\vspace{-0.5 cm}
\end{figure*}

To understand the specifics of our implementation, we examine sorting from an operational point of view. Sorting is a unitary operation that bijectively maps input photons of different modes onto different output modes. One approach to realize such an operation is by successive application of a discrete Fourier transform (i.e.\,  $F$-gate), a mode-dependent phase unit (i.e.\, $Z_d$-gate), and an inverse discrete Fourier transform element \cite{ionicioiu2016sorting} {(Note that we use the quantum gates and the bracket notation in order to provide a concise mathematical description for the evolution of spatial modes, and not for the purpose of describing the quantum state of the electromagnetic field).} The discrete Fourier transform can be realized by a combination of beam-splitters and constant-phase elements (wave-plates) \cite{PhysRevLett.73.58, tabia2016recursive}. The remaining unit required for sorting the LG modes according to their radial index is a mode-dependent phase element i.e. a $Z_d$-gate.

We next describe how the $Z_d$-gates for the LG modes can be realized using a natural property of these modes in propagation. The mathematical form of the LG modes in cylindrical coordinates at the plane of the beam waist is given by \cite{plick2015physical}:
\begin{align}\nonumber
\text{LG}_{\textit{p} \ell}(  r,\theta)= & \sqrt{\frac{2p!}{\pi (p+|\ell |)!}}  \frac{1}{w_0} \left(\frac{\sqrt{2}r}{w_0}\right)^{|\ell|}  \\
 & \times  \text{exp}\left( -\frac{r^2}{w_0^2}\right) L_p^{|\ell|}\left(\frac{2r^2}{w_0^2}\right) e^{i\ell \theta}
\end{align}
where $L_p^{|\ell|}$ is the generalized Laguerre polynomial and $w_0$ is the beam waist radius. It is a well-known fact that these modes are eigenmodes of a family of linear transforms generalizing the Fourier transform. This family of operations  are the Fractional Fourier transforms (FRFTs), and the characteristic equation for LG modes is given by \cite{Supplement, HankelNote}
\begin{equation}\label{Eq:FGphase}
\mathcal{F}^{a } [\text{LG}_{p\ell}(r_0,\theta_0) ] =\text{exp}[-i(2p+|\ell|)  {a}] \text{LG}_{p\ell}(r,\theta)
\end{equation}
In the above equation $a$ denotes the order of the FRFT and for a normal Fourier transform it is $\pi/2$. The phase term here can be interpreted as a modification of the effective phase velocity of the structured beam, and is reminiscent of the Gouy phase in laser physics. For the purposes of this paper we refer to this mode-dependent phase as the fractional Gouy phase \cite{Kogelnik:1966ik,yew2008fractional}.

\begin{figure*}[t!]
\includegraphics[width= 0.8\textwidth]{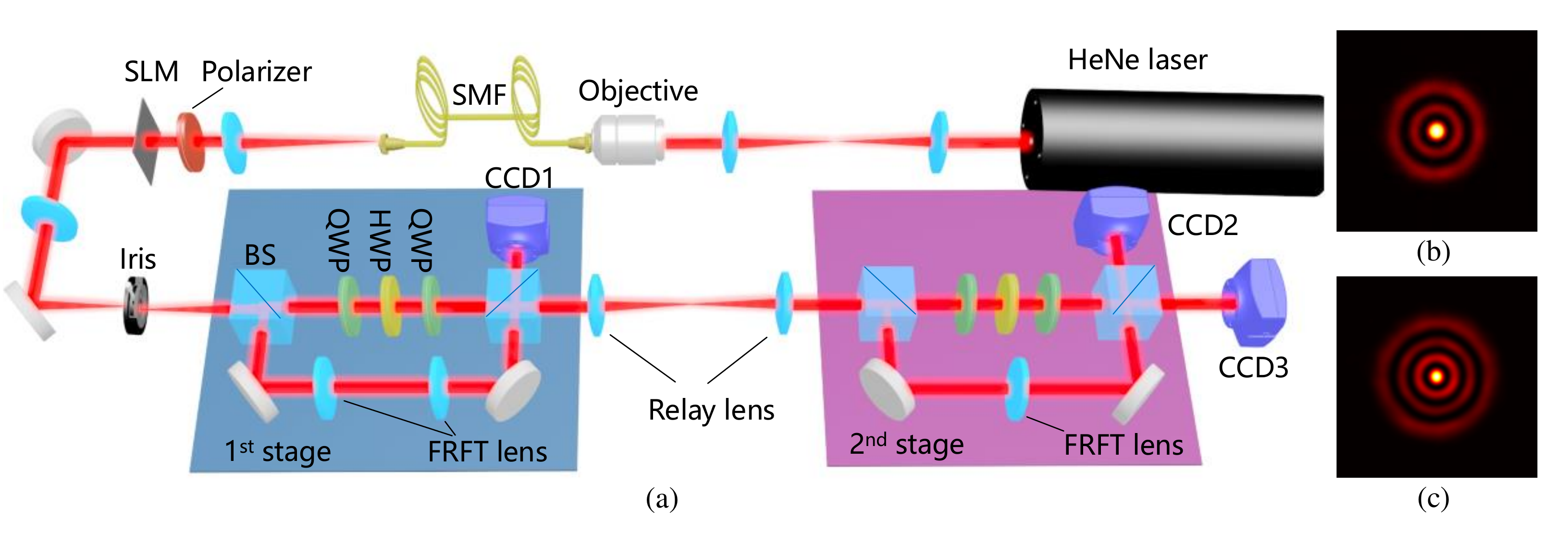}
\caption{
(a) Schematics of the experimental setup. The radial mode is generated by a computer generated hologram (CGH) on the spatial light modulator (SLM). The quarter waveplate (QWP), half waveplate (HWP) and QWP combination works as a geometrical phase shifter. Further detail about the setup can be found in the Supplementary Material \cite{Supplement}. (b, c) The measured intensity profile of the generated $p=2$ and the $p=3$ modes.}
\label{fig:setup}
\vspace{-0.5 cm}
\end{figure*}

A simple operational unit of our mode sorter, consisting of a single lens accompanied with free-space propagation, can realize the FRFT (see Fig.~\ref{fig:FRFT} (a)) \cite{lohmann1993image}. The propagation distance $z$ and the lens focal length $f$ are related to the FRFT order $a$, the wavelength $\lambda$, and the beam waist radius $w_0$ through the following equations \cite{Supplement}:
\begin{align}\label{Eq:parameters}
z = \frac{\pi w_0^2}{\lambda}\tan\frac{a}{2}, ~~~~~~~~~
f = \frac{\pi w_0^2}{\lambda \sin a}
\end{align}
Upon propagation through this unit, radial modes will pick up a fractional Gouy phase that depends on their respective indices. Note that the corresponding phase depends on both the radial index as well as the OAM value. This dependence does not present a problem as one can use a Dove prism to cancel the $\ell$-dependence and thereby retain only the $p$-dependent phase \cite{leach2002measuring}.

Having examined the two building blocks, i.e. the discrete Fourier transform ($F$-gate) and the $Z_d$-gate, we can design a radial index mode sorter. A schematic representation of the concept is provided in Fig.~\ref{fig:FRFT} (b). Let us assume that $\ell=0$ and denote the LG mode by $\ket{p}$. We suppose that the dimension of the state space is $d$, and that $p$ takes on the values $0, 1, ..., d-1$. The output port for each mode is represented by a different ket $\ket{k}$, where $k=0, 1, ..., d-1$.
Initially, all modes are present in the same input port $ \ket{k_{in}}$, and the state vector is denoted by $\ket{p}\otimes \ket{k_{in}}$. To sort different modes according to their radial indices, we ensure that their output ports depend only on their radial indices. This operation can be expressed as $\ket{p}\otimes \ket{k_{in}} \mapsto \ket{p}\otimes \ket{k=p}$. The successive application of a discrete Fourier transform ($F$-gate), a $Z_d$-gate, {and a $F^{\dagger}$-gate can realize this transformation.} The explicit transformation that each gate provides is given below:
\begin{equation}
\begin{aligned}
&\hat{F}\Big[ \ket{p}\otimes \ket{k}  \Big]=\frac{1}{\sqrt{d}}\sum_{m=0}^{d-1} \text{exp}\left(\frac{i2\pi mk}{d}\right)\ket{p}\otimes \ket{m} \\
&\hat{Z}^j_d \Big[  \ket{p}\otimes \ket{k} \Big] =\text{exp}\left(\frac{i2\pi pj}{d}\right)  \ket{p}\otimes \ket{k}.
\label{eq:Fgate}
\end{aligned}
\end{equation}
where $\hat{F}$ and $ \hat{Z}_d$ indicates the $F$- and $Z_d$-gate respectively, and $j$ is the order of the corresponding $Z_d$-gate. {The $F^{\dagger}$-gate is the inverse $F$-gate.} A $Z_d$-gate of order $j$ is equivalent to $j$ subsequent applications of the $Z^{1}_d$-gate \cite{almeida1994fractional}.

In the first part of our implementation, we realize a binary version of our proposed radial sorter. By setting $d=2$ in Eq. (\ref{eq:Fgate}), the setup reduces to an interferometer with a FRFT in one of the arms. To have more control over the phase we also include a constant phase shifter in the other arm. The $Z_d$-gate unit introduces a fractional Gouy phase to each of the input modes and causes distinct input modes to interfere constructively at different output ports. Thus photons of different radial indices leave the interferometer at different output ports and the sorting transformation is achieved. We note that \citet{leach2002measuring} have previously demonstrated a conceptually similar design for an OAM mode sorter.

In the next step, we increase the dimensionality of the system by cascading two successive binary sorters of the type shown in Fig.~\ref{fig:FRFT}(b). This configuration allows us to sort up to three radial modes. Compared to the multi-channel interferometer proposed in \cite{ionicioiu2016sorting}, this cascading scheme is advantageous in terms of flexibility, complexity and practicality (For a comparative analysis please refer to the Supplementary Material \cite{Supplement}.) A schematic representation of our setup is depicted in Fig.~\ref{fig:setup}. A 633 nm HeNe laser is coupled to a single mode fiber (SMF). The light emerging from the fiber is then collimated to illuminate a spatial light modulator (SLM). A binary computer generated hologram is imprinted onto the SLM to generate the desired field in the first diffraction order \cite{Davis:1999ku,mirhosseini2013rapid}. In the first stage, we use a lens with a focal length of \mbox{30 cm} and with a propagation distance of $z=8.79 \text{ cm}$ to realize a FRFT of the order $\pi/4$ for a beam waist radius of $w_0= $ \SI{207}{\micro\meter}. The second stage of the sorter uses two lenses with the same configuration to provide a FRFT with twice as much phase shift. We note that the interferometer shown in the schematic is imbalanced because of the need to introduce the FRFT lenses in one arm. We have taken care to keep the path imbalance much shorter than the coherence length of our laser source and the Rayleigh ranges of our modes.

\begin{figure}[t!]
\includegraphics[width=\columnwidth]{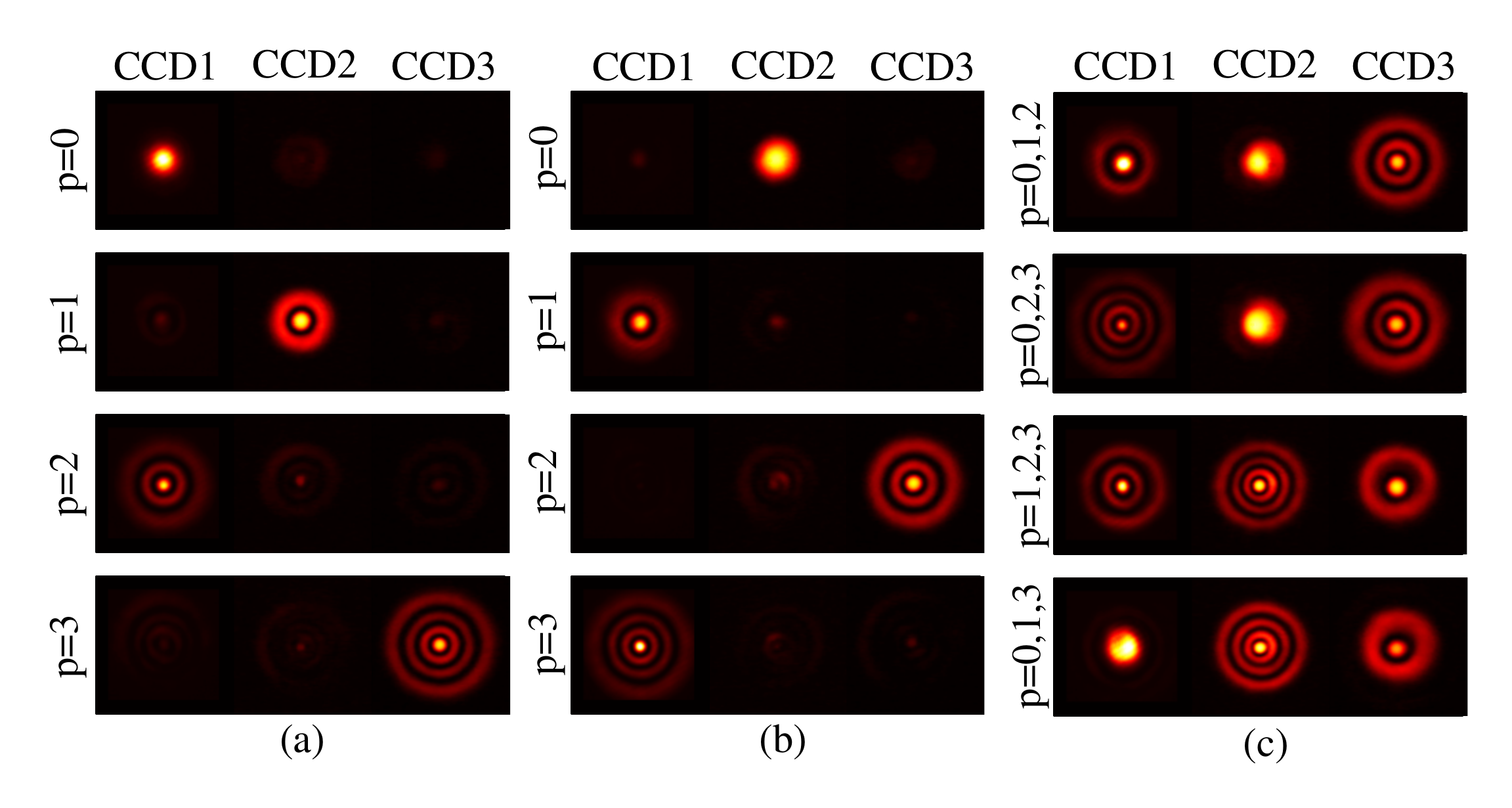}
\caption{
Output port image for inputs in the form of individual LG modes and their superposition states. The position of each CCD is shown in Fig.~\ref{fig:setup}. (a) The path lengths in the first stage are adjusted so the even-order modes are sent to CCD1 whereas the odd-order modes are sent to the second stage where they are further sorted so that $p=1$ ($p=3$) is directed to CCD2 (CCD3). (b) The phase shifter in the first stage is readjusted to send odd-order modes to CCD1 and the even-order modes to the second stage. (c) The images on CCDs when a superposition state is sent to the sorter. $p=0,1,2$ means that a superposition state composed of $p=0$, $p=1$ and $p=2$ mode is generated and injected. All images in the same line are captured simultaneously.}
\label{fig:fullresult}
\vspace{-0.5 cm}
\end{figure}

In order to characterize the proposed scheme, we first generate radial modes and detect the output of our setup using charge coupled devices (CCD). The images from the three CCD cameras at the three output ports of the setup are shown in Fig.~\ref{fig:fullresult}(a) and (b). In Fig.~\ref{fig:fullresult}(a) even-order modes $(p=0,2)$ leave one of the output ports of the first binary sorter to CCD1 and the odd-order modes leave the other output port. The odd-order modes are then fed into the second stage, and are routed towards CCD2, and CCD3. By changing the phase in the first stage one can send odd-order modes to CCD1 and send the even-order modes to the second stage to be sorted to CCD2, and CCD3. The cascaded binary sorters allow for sorting of up to three separate modes. As an additional test of the validity of our scheme we produce linear superpositions of three radial modes and feed it into the first stage. We then register the image of the three output ports on the CCDs simultaneously. It is clear from Fig.\,\ref{fig:fullresult}(c) that although all the input photons share a superposition of three radial indices, the output photons are sorted according to their radial indices. We note that to sort different sets of modes one has to choose appropriate phase differences for the two binary sorters. The value of the induced phases are different for two different sets of modes, and can be calculated using the formula for the fractional Gouy phase in Eq\,\ref{Eq:FGphase}. Indeed, \textit{a priori} knowledge about the input state is necessary for an appropriate sorting. For any finite-dimensional sorter the input state should be restricted to a specific range.

\begin{figure}[t!]
\includegraphics[width=\columnwidth]{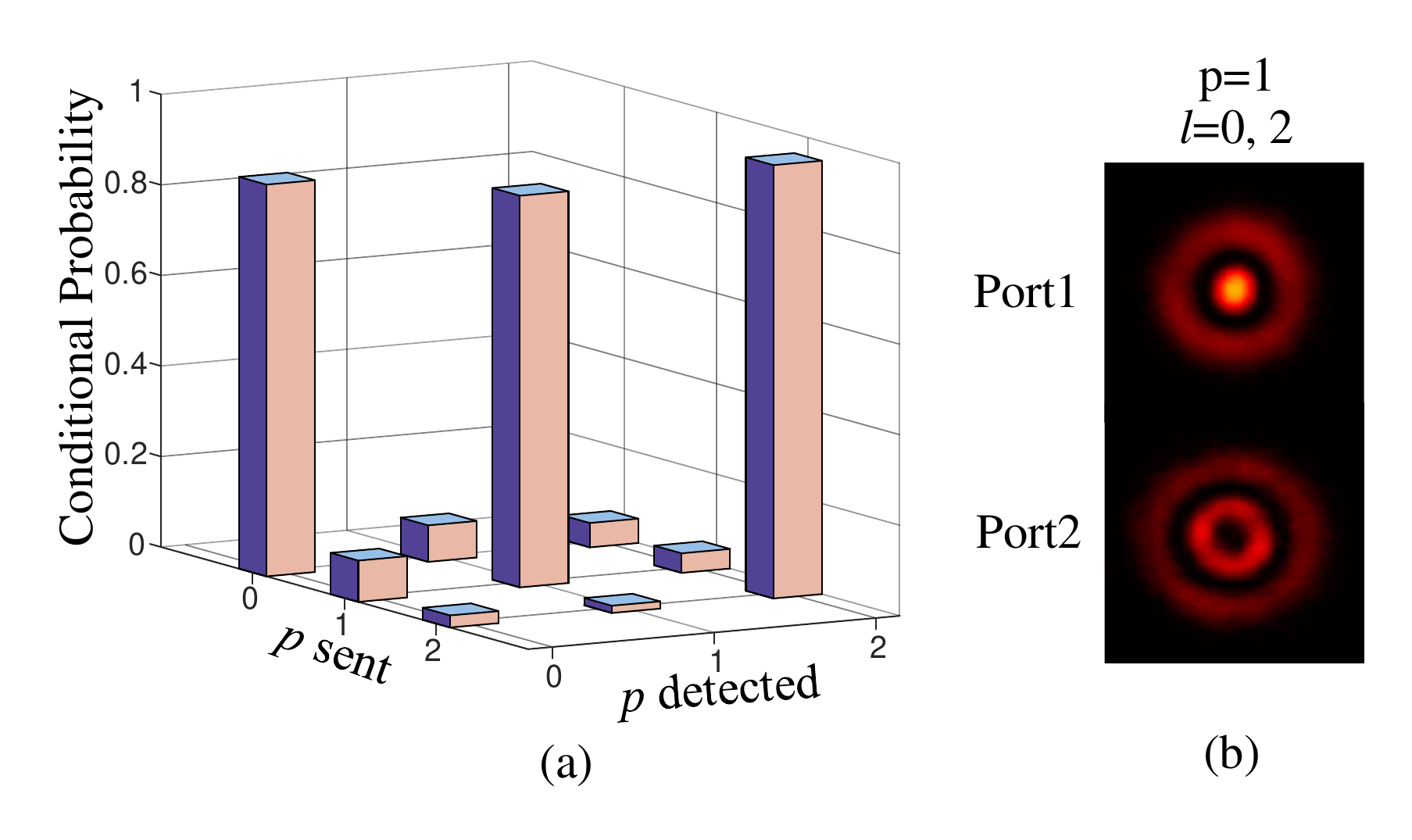}
\caption{(a) Experimentally measured probability of detection as a function of input and output mode indices. (b) The measured output intensity profile for an input prepared as a superposition of $p=1, \ell=0$ and $p=1, \ell=2$ modes.}
\label{fig:crosstalk}
\vspace{-0.5 cm}
\end{figure}

As mentioned above, our scheme can also be used for sorting of photons according to their OAM number.  To demonstrate this capability we use the first stage of our setup to implement a binary sorting of LG$_{10}$ and LG$_{12}$. The images of the output ports are plotted in Fig.\,\ref{fig:crosstalk}, and confirm that photons of different OAMs leave the interferometer at separate ports. We underscore the fact that here we have separated two OAM modes of the same radial order whose OAM values are different by $\Delta \ell=2$. The spacing by two units results from the fact that the phase shift from the FRFT is $\Delta \phi=(2p+\ell)a$. The extra factor 2 for $p$ index implies that the $\ell$ spacing has to be twice larger. Of course, by selecting the appropriate order of the FRFT, our device can sort the LG beams with $\Delta \ell=1$ as well.

We have quantified the cross-talk of our setup by measuring the conditional probability matrix. Each element of this matrix is defined as the probability of detecting a photon at a given mode conditioned on the radial index of the input. This quantity is equal to the power in a specific port divided by the total output power. The resulting matrix is plotted in Fig.~\ref{fig:crosstalk}. To use a single figure of merit we use the total cross-talk, which is sum of the power in the wrong ports divided by the total output power. For our specific implementation the total cross-talk is measured to be 15\%. In addition we wish to emphasize that this cross-talk is not intrinsic to the protocol. We believe that using high-quality anti-reflection coated optics, active stabilization, and more careful alignment can mitigate cross-talk significantly and bring the sorter to its theoretical limit of 100\% efficiency and no cross-talk.

We note that our design can also be employed for sorting the Hermite-Gaussian (HG) modes. Coherent detection of LG and HG modes has been recently identified as an optimal means of localizing closely spaced incoherent sources \cite{tsang2016quantum, paur2016achieving,PhysRevLett.118.070801,yang2016far}. It is thus reasonable to expect that an efficient sorting mechanism can have further implications for microscopy, given the significance of super-resolution in that field. In addition, a similar approach can be applied to sorting the family of Bessel-vortex beams. Due to the non-diffracting property of these modes, free-space propagation can serve as the $Z$-gate and there is no need for realization of the FRFT module. Hence, a simplified version of our experiment with the the FRFT components removed would be able to sort Bessel beams with different longitudinal wavevectors.

In summary we have demonstrated a general framework for efficient measurement (i.e.\,sorting) of the radial index of LG modes. Our protocol includes two essential elements: the discrete Fourier transform ($F$-gate) and the $Z_d$-gate. While discrete Fourier transform can be realized using beam splitters and wave-plates, we have employed the fractional Gouy phase to realize the $Z_d$-gate efficiently. As a demonstration we have implemented a binary $(d=2)$ version of our protocol and have cascaded two binary sorters to sort three different LG modes according to their radial indices. {Combined with a total angular momentum sorter, our protocol provides a platform for accessing the complete spatial bandwidth of photons for encoding information. We believe that implementation of our protocol can facilitate fundamental studies of the spatial modes of light as well as a variety of prevalent applications of such states in quantum communications,  imaging, and quantum metrology \cite{nielsen2010quantum}}.

\begin{acknowledgments}
The authors thank Jerry Kuper for technical support and Boshen Gao for helpful discussion. This work is supported by the US Office of Naval Research. In addition, RWB acknowledges support from Canada Excellence Research Chairs Program and National Science and Engineering Research Council of Canada. DF acknowledges support from National Natural Science Foundation of China (Grant Nos. 11374008 and 11534008) and China Scholarship Council overseas scholarship.
\end{acknowledgments}

%

\pagebreak
\widetext
\begin{center}
\textbf{\large Supplemental Materials: sorting photons by radial quantum number}
\end{center}

\setcounter{equation}{0}
\setcounter{figure}{0}
\setcounter{table}{0}
\setcounter{page}{1}
\makeatletter
\renewcommand{\theequation}{S\arabic{equation}}
\renewcommand{\thefigure}{S\arabic{figure}}
\renewcommand{\bibnumfmt}[1]{[S#1]}
\renewcommand{\citenumfont}[1]{S#1}

\section{ Mathematical derivations}
In the primary article, we have stated that the Laguerre-Gaussian (LG) function is the eigenfunction of the fractional Fourier transform (FRFT) with eigenvalue $\text{exp}[-i(2p+ |\ell |)\pi]$. We mention that the beam size of the LG modes should satisfy $\lambda \tilde{f}/ \pi =w_0^2$ in order to remain invariant after the FRFT.

\subsection{Eigenvalue of the Laguerre-Gaussian funciton}

In this section we derive an expression for the eigenvalue of the LG function for the FRFT, which can be expressed as
\begin{equation}
\mathcal{F}^{a } [\text{LG}_{p \ell }(r_0,\theta_0) ] =\text{exp}[-i(2p+| \ell |)a] \text{LG}_{p \ell }(r,\theta)
\label{eq:eigenvalue}
\end{equation}
The proof of this expression can be attained by using the relation between the LG function and the Hermite-Gaussian (HG) function. It is well known that the LG function can be decomposed to the HG function as follows \cite{Sbeijersbergen1993astigmatic}
\begin{equation}
\text{LG}_{p \ell }(r,\theta) = \sum_{k=0}^{N=m+n}i^kb_{n,m,k}  \text{HG}_{N-k,k}(x,y)
\label{eq:decompose}
\end{equation}
where $b_{n,m,k}$ is a constant determined by $n, m \text{ and } k$, and its definition is
\begin{equation}
b_{n,m,k}=\sqrt{\frac{(N-k)!k!}{2^Nn!m!}}\frac{1}{k!}\frac{d^k}{dt^k}  \left[   (1-t)^n(1+t)^m  \right]  _{t=0}
\end{equation}
where the relation between indices $m$, $n$ and $p$, $\ell$ is given by $p=\text{min}(m,n),\ell = m-n$ and $2p+ |\ell |=m+n$. Since the HG function is the eigenfunction of the FRFT \cite{Salmeida1994fractional}
\begin{equation}
\mathcal{F}^{a }[ \text{HG}_{mn}(x_0,y_0) ] =\text{exp}[-i(m+n)a]\text{HG}_{mn}(x,y)
\label{eq:HGeigen}
\end{equation}
we can use Eq.~(\ref{eq:decompose}) to decompose the LG function to the HG function, and then transform the resulting HG functions through Eq.~(\ref{eq:HGeigen}). Then we can derive the transformed LG function, which can be expressed as
\begin{equation}
	\begin{aligned}
 \mathcal{F}^{a } &[\text{LG}_{p \ell }(r_0,\theta_0) ] \\
&=  \mathcal{F}^{a } \left[   \sum_{k=0}^{N=m+n}i^kb_{n,m,k}  \text{HG}_{N-k,k}(x_0,y_0)  \right] \\
 &=\sum_{k=0}^{N=m+n}i^kb_{n,m,k}  \mathcal{F}^{a }[\text{HG}_{N-k,k}(x_0,y_0)]
	 \\
 &=\sum_{k=0}^{N=m+n}i^kb_{n,m,k}  \text{ exp}(-iNa) \cdot \text{HG}_{N-k,k}(x,y)
	\\
    &=\text{exp}[-i(2p+| \ell |)a]\text{LG}_{pl}(r,\theta)\\
	\end{aligned}
\end{equation}
which is the proof to Eq.~(\ref{eq:eigenvalue}).

\subsection{Beam waist radius of the Laguerre-Gaussian mode}
Previously we have mentioned that a single lens can perform the FRFT along with a free space propagation. However, only the LG beam of a specific beam waist radius can remain invariant after the FRFT. The beam waist radius should satisfy the condition $\lambda \tilde{f}/ \pi =w_0^2$, and the derivation is presented as follows.

Assume the optical field on the initial plane is $u_0(x_0,y_0)$ and the field on the final plane is $u(x,y)$. The field propagates a distance of $z$, goes through a lens of focal length $f$, and propagates another $z$ again. By Fresnel propagation, the relation between $u_0(x_0,y_0)$ and $u(x,y)$ is \cite{Sgoodman2005introduction}
\begin{equation}
\begin{aligned}
&u(x,y)  \propto \text{exp}\left[\frac{i\pi}{\lambda \tilde{f} \text{tan }a}\left(x ^2+y ^2\right)\right] \iint dx_0 dy_0 u_0(x_0,y_0) \\
& \times \text{exp}  \left[  \frac{i\pi}{\lambda \tilde{f} \text{tan }a}\left(x_0^2+y_0^2\right)  \right]  \text{exp}\left[-\frac{ 2i\pi}{\lambda \tilde{f} \text{sin }a} \left(  xx_0+yy_0\right)  \right]
\label{eq:FRFTphysical}
 \end{aligned}
\end{equation}
where $f=\tilde{f}/\text{sin }a$, $z=\tilde{f}\text{tan}(a/2)$ and we have omitted the normalization factor. By defining the variables $X=x \big/ \sqrt{\lambda \tilde{f}/2\pi}$, $Y=y \big/  \sqrt{\lambda \tilde{f}/2\pi}$, $X_0=x_0  \big/   \sqrt{\lambda \tilde{f}/2\pi}$ and $Y_0=y_0  \big/   \sqrt{\lambda \tilde{f}/2\pi}$, we can rewrite the equation and express it as
\begin{equation}
\begin{aligned}
&u(x,y) \propto  \text{exp}\left[\frac{i}{2\text{tan }a}\left(X ^2+Y ^2\right)\right]  \iint \text{exp}\left[\frac{i\left(X_0^2+Y_0^2\right)}{2\text{tan }a}\right]\\
& \times u_0 \left(   \sqrt{\frac{\lambda \tilde{f}}{2\pi}}X_0,\sqrt{\frac{\lambda \tilde{f}}{2\pi}}Y_0 \right)\text{exp}\left[-\frac{ i(XX_0+YY_0)}{ \text{sin }a}\right]dX_0 dY_0
 \end{aligned}
\end{equation}
Comparing to definition of the two-dimensional FRFT \cite{Salmeida1994fractional}
\begin{equation}
\begin{aligned}
 \mathcal{F}^{a}&[u_0(x_0,y_0)] =\frac{1-i\text{cot }a}{2\pi} \iint dx_0dy_0 u_0(x_0,y_0)\\
 &\times \text{exp}\left[i \left(\frac{x^2+y^2}{2\text{tan }a}- \frac{xx_0+yy_0}{\text{sin }a}+\frac{x_0^2+y_0^2}{2\text{tan }a}\right)\right]
  \label{eq:FRFT}
\end{aligned}
\end{equation}
we could readily check that $u(x,y)$ can be represented in the form of the FRFT as
\begin{equation}
\begin{aligned}
u(x,y) &\propto    \mathcal{F}^a_{X_0\rightarrow X, Y_0\rightarrow Y} \left[u_0\left(\sqrt{\frac{\lambda \tilde{f}}{2\pi}}X_0,   \sqrt{\frac{\lambda \tilde{f}}{2\pi}}  Y_0\right)\right]
 \end{aligned}
\end{equation}
The subscript $X_0\rightarrow X, Y_0\rightarrow Y$ means that the FRFT is mapping the new variables $X_0,Y_0$ to $X\text{ and }Y$, respectively, not acting on the original $x$ and $y$. Now assume the incident beam is a LG beam of the beam waist radius $w_0$. So we can express it as
\begin{equation}
\begin{aligned}
&u_0( x_0,y_0) = \text{LG}_{p\ell } \left(\frac{r_0}{w_0/\sqrt{2}}, \theta \right) \propto   \frac{1}{w_0} \left(\frac{r_0}{w_0/\sqrt{2}}\right)^{|\ell |} \\
& \times \text{exp}\left[-\frac{1}{2} \left(\frac{r_0}{w_0/\sqrt{2}}\right)^2\right]L_p^{|\ell |}\left(\left(\frac{r_0}{w_0/\sqrt{2}}\right)^2\right)\text{exp}(i \ell \theta)
 \end{aligned}
\end{equation}
Here we ignore the normalization factor. We can play the same trick to decompose the LG function to the HG function as
\begin{equation}
\begin{aligned}
& u_0(x_0, y_0) =
\text{LG}_{p\ell }\left(\frac{r_0}{w_0/\sqrt{2}}, \theta \right) \\
&=  \sum_{k=0}^{N=m+n}i^kb_{n,m,k}\text{HG}_{N-k,k}\left( \frac{x_0}{w_0/\sqrt{2}}, \frac{y_0}{w_0/\sqrt{2}} \right) \\
&=  \sum_{k=0}^{N=m+n}i^kb_{n,m,k}\text{HG}_{N-k,k}\left( \frac{\sqrt{\lambda \tilde{f}/2\pi}}{w_0/\sqrt{2}}X_0, \frac{\sqrt{\lambda \tilde{f}/2\pi}}{w_0/\sqrt{2}}Y_0 \right) \\
&=  \sum_{k=0}^{N=m+n}i^kb_{n,m,k}\text{HG}_{N-k,k}
\left( \frac{\sqrt{\lambda \tilde{f}/ \pi}}{w_0 }X_0, \frac{\sqrt{\lambda \tilde{f}/ \pi}}{w_0}Y_0 \right) \\
 \end{aligned}
\end{equation}
where $p=\text{min}(m,n)$ and $\ell =m-n$. Clearly, if we have $ \sqrt{  \lambda \tilde{f} /   \pi w_0^2 }=1$, or equivalently $\lambda \tilde{f}/ \pi =w_0^2$, then the field on the final plane can be simplified to
\begin{equation}
\begin{aligned}
&u(x,y)  \propto  \sum_{k=0}^{N=m+n}i^kb_{n,m,k}  \mathcal{F}^a_{X_0\rightarrow X, Y_0\rightarrow Y}  [\text{HG}_{N-k,k}( X_0 , Y_0  )  ]\\ &=\sum_{k=0}^{N=m+n}i^kb_{n,m,k}  \text{exp}(-iNa)\text{HG}_{N-k,k}( X , Y )   \\
& = \text{exp}(-iNa) \sum_{k=0}^{N=m+n}i^kb_{n,m,k} \text{HG}_{N-k,k}\left( \frac{x}{w_0/\sqrt{2}} , \frac{y}{w_0/\sqrt{2}} \right)\\
&=\text{exp}[-i(2p+|\ell |)a]\text{LG}_{p \ell }\left(\frac{r}{w_0/\sqrt{2}},\theta\right)
 \end{aligned}
\end{equation}
We could readily see that the field on the final plane is a LG beam with the same beam waist radius $w_0$ as long as the condition $\lambda \tilde{f}/ \pi =w_0^2$ is satisfied. In our experiment, we choose the order of the FRFT to be $a=\pi/4$, and the lens focal length to be 30 cm. Using this condition, the beam waist radius is calculated to be $w_0= $ \SI{207}{\micro\meter}.

\subsection{Comparison between the cascading and multi-channel interferometer}

In this section, we compare the cascading interferometer presented in the manuscript and the multi-channel interferometer proposed in Ref.~\cite{Sionicioiu2016sorting} in terms of complexity, flexibility and practicality.

(1) A big advantage of the cascaded scheme is that it has a simpler structure for high dimensional system. In the multi-channel interferometer shown in the Fig.~1 in the primary article, there are two quantum $F$-gates which can be implemented by the 50/50 beam splitters. For a $2^n$-dimensional sorter, the first quantum $F$-gate is effectively a 1-to-$2^n$ splitter and it requires $2^n-1$ beam splitters. The second quantum $F$-gate, however, needs $(n+2)\cdot 2^{n-1}-1$ beam splitters and the calculation is as follows. Assume a $d$-dimensional $F$-gate requires $N(F_{d})$ beam splitters. Then we will have the following recursive relation \cite{Stabia2016recursive}
\begin{equation}
\begin{aligned}
N(F_{2d}) = 2\cdot N(F_d) + d
 \end{aligned}
\end{equation}
With the knowledge that $N(F_2)=1$, we can arrive at the following equation $N(F_d)=(d/2)\cdot \text{log}_2d$. Setting $d=2^n$, we will have $N(F_{d=2^n})=n\cdot 2^{n-1}$. So the total number of beam splitters is $n\cdot 2^{n-1} + 2^n - 1=(n+2)\cdot 2^{n-1}-1$ for the multi-channel interferometer.

As for the cascading interferometer presented in the primary article, we will need $2^n-1$ Mach-Zenhder interferometer (MZI) to sort $2^n$ modes, and each MZI requires two beam splitters. So the total number is $2\cdot (2^n-1)= 2^{n+1}-2$. It can be readily verified that $ 2^{n+1}-2 <(n+2)\cdot 2^{n-1}-1 $ for $n \geq 2$ .

(2) A multi-channel interferometer requires that all paths remain in-phase simultaneously. This can cause a serious challenge for experimental realization as any phase mismatch will compromise the functionality of all output channels. In contrast, if one of the sorter paths in the cascading scheme is not in-phase, the system continues to properly operate for the channels that do not utilize the mismatched path.

(3) The conventional quantum $F$-gate is usually composed of 50/50 beam splitters and phase shifters \cite{Stabia2016recursive}, and thus it always has a dimensionality of $d=2^n$, where $n$ is an integer. Thus it would be difficult to realize a sorter with a dimensionality that is not a power of 2.  Taking the first quantum $F$-gate as an example, one would need to use 33/67 beam splitters to build a $d=3$ system, and 20/80 beam splitters for $d=5$ system. Considering this, the cascaded approach is simpler to realize due to the commercial availability of 50/50 beam splitters.

\section{Experimental details}
In this section we describe the experimental details of our setup. We use a combination of waveplates as a geometrical phase shifter, which can provide a continuously adjustable control on the phase of the travelling beam. Also the implementation of the two-dimensional quantum $F$-gate by the polarizing beam splitter (PBS) and half waveplate (HWP) is discussed. We also describe how we generate the LG modes used in this experiment.
\begin{figure}[t]
\centering
\includegraphics[width=0.5\linewidth]{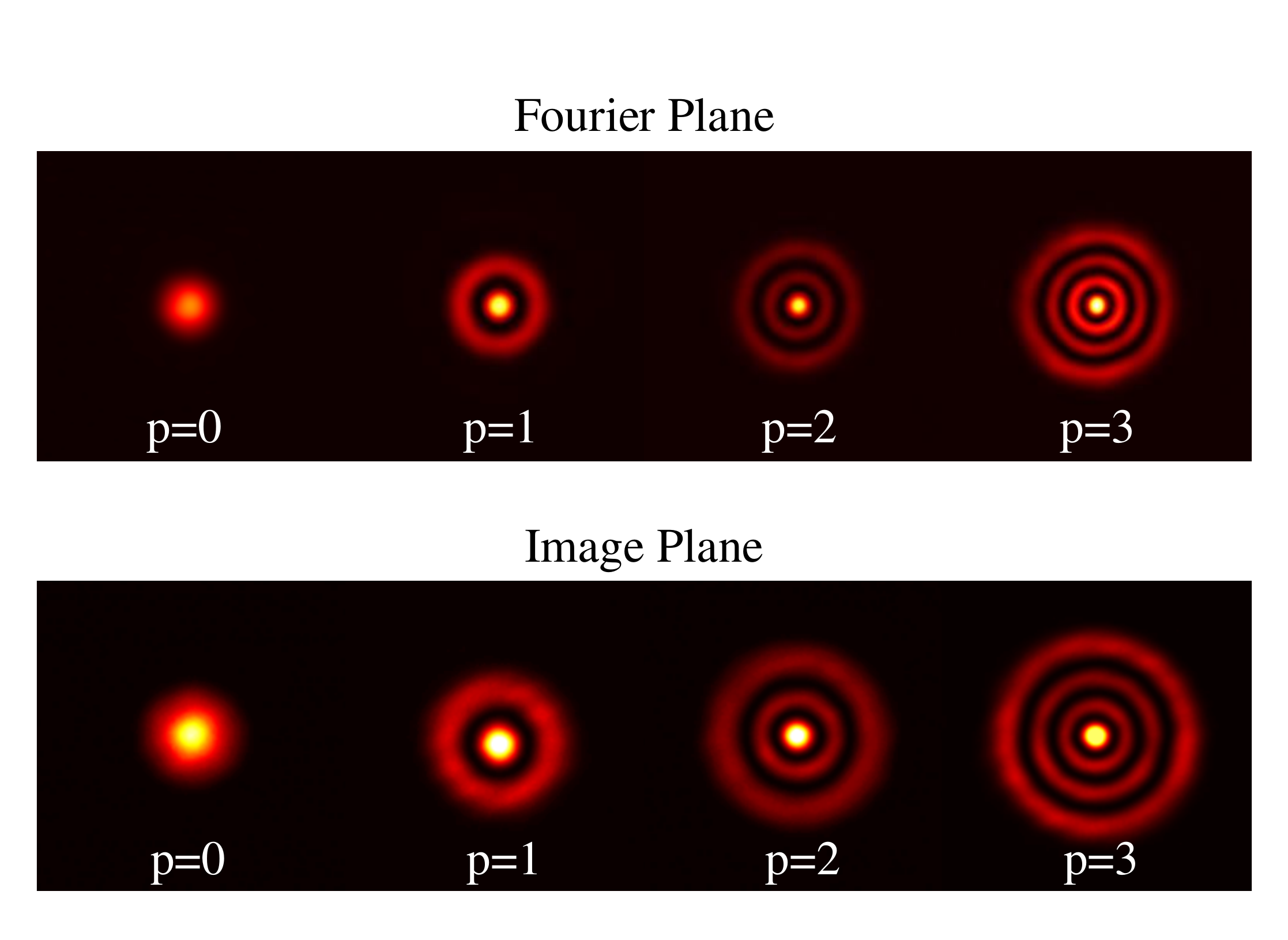}
\caption{Generated radial modes in the Fourier plane and image plane. }
\label{fig:Modes}
\end{figure}

\subsection{Geometrical phase shifter}
In our setup the SLM only modulates the horizontally polarized light, thus the prepared state is horizontally polarized automatically, and the polarization state vector can be expressed as $\textbf{ P} =[1 \text{ }0]^T$, where the superscript $T$ means the transpose of a vector. The geometrical phase shifter is composed of two quarter waveplates (QWP) and a HWP. The two QWPs are $45^{\circ}$ oriented while the HWP between them has an angle of $\theta$. The Jones matrix of the geometrical phase shifter is calculated to be
\begin{equation}
\begin{split}
    \textbf{ M} &=\frac{\sqrt{2}}{2} \begin{bmatrix}
      i & 1 \\1 & i
     \end{bmatrix}\cdot
    \begin{bmatrix}
        \text{cos }2\theta & \text{sin }2\theta \\ \text{sin }2\theta & -\text{cos }2\theta
    \end{bmatrix} \cdot \frac{\sqrt{2}}{2}
        \begin{bmatrix}
       i & 1 \\1 & i
     \end{bmatrix} \\     &=\begin{bmatrix}
     -e^{-i2\theta} & 0 \\ 0 &e^{i2\theta}
     \end{bmatrix}
\end{split}
\end{equation}
 Going through the waveplates will lead the state vector to be $\textbf{ P}' =\textbf{ M} \cdot [1 \text{ }0]^T=-e^{-i2\theta}[1 \text{ }0]^T$. Hence, by adjusting the angle of the HWP, we can induce an arbitrary phase shift, which is equivalent to the piezoelectric actuator. We emphasize that our proposed sorting method is intrinsically polarization-independent, and the PBS and HWP can be replaced by the non-polarizing beam splitter (NPBS) while the geometrical phase shifter can be replaced by a piezoelectric actuator for a polarization-independent application.

\subsection{Two-dimensional quantum $F$-gate}
The two-dimensional quantum $F$-gate performs the following transform
\begin{equation}
\hat{F}\Big[ \ket{p}\otimes \ket{k}\Big]=\frac{1}{\sqrt{2}} ( \ket{p} \otimes \ket{0} + \text{exp}( i  \pi ) \ket{p} \otimes \ket{1})
\end{equation}

As can be seen, this two-dimensional quantum $F$-gate will split the mode into two copies with an extra phase on one of them. Since we can always adjust the phase by the geometrical phase shifter, by applying a non-polarizing beam splitter (NPBS) we can effectively form this $F$-gate. Hence, the two-dimensional sorter which consists of two quantum $F$-gates and a quantum Z-gate has a structure similar to the Mach-Zehnder interferometer. However, the NPBS should possess a 50/50 splitting ratio, which is not always true for the commercially available broadband NPBS. Here we replace the first NPBS by a HWP and a PBS. Remember that the incident beam is automatically horizontally polarized due to the SLM, so we can always rotate the polarization angle and control the splitting ratio. After a HWP which can rotate the polarization by $45^{\circ}$, the two beams split by the first PBS are horizontally and vertically polarized, respectively. The second PBS will recombine the two beams after the geometrical phase shifter. However, they cannot interfere due to their orthogonal polarizations. Then we can again apply a HWP to rotate two beams by $45^{\circ}$ and then split the beam by another PBS, which forms the second $F$-gate. Now the beam in each output port will have the same polarization and can interfere effectively. By aligning the rotation angle of the HWP, we can reach an arbitrary splitting ratio and can also help compensate the possible unequal loss in two paths.

\subsection{Radial mode generation}
To prepare the radial modes, we imprint the computer generated hologram on a spatial light modulator (SLM) \cite{Smirhosseini2013rapid}. The binary phase grating will generate the mode at the first diffraction order, which can be separated by using a Fourier-transforming lens. To verify the quality of generated mode, it is necessary to check the mode on not only the Fourier plane but also the image plane (which can accessed by performing  a second successive Fourier transformation). The generated mode on two planes by our setup are presented in Fig.~\ref{fig:Modes}. We note that the polarization state of the light beam on the SLM should be aligned according to the SLM requirement in order to minimize interference with the strong zeroth-order light. We use a 1500 mm lens along with an iris to separate the first order diffracted beam. In order to increase the fidelity of the generated beam, we have applied an additional phase term to the computer generated hologram to correct for spherical aberration, astigmatism and the coma, which are typically caused by the imperfections in the system.

\end{document}